\documentclass[12pt]{article}

\usepackage[letterpaper,margin=3cm]{geometry}

\usepackage{authblk}

\usepackage[parfill]{parskip}
\usepackage{setspace}

\usepackage[labelfont=bf,font={small,stretch=0.95}]{caption}

\usepackage[super,sort,compress,numbers]{natbib}

\usepackage{gensymb}
\usepackage{amsmath}
\usepackage{amssymb}
\usepackage{mathtools}
\usepackage{comment}
\usepackage{hyperref}

\usepackage{graphicx}
\usepackage{multirow}

\usepackage[version=4]{mhchem}

\usepackage{mathptmx}

\usepackage{xcolor}
\usepackage{soul}
\usepackage{booktabs} 
\usepackage[utf8]{inputenc}
\usepackage[T1]{fontenc}
\usepackage{lmodern}

\def\go{\texttt{GO-MACE-23}}
\def\off{\texttt{MACE-OFF}}
\def\offt{\texttt{MACE-OFF23}}
\def\offf{\texttt{MACE-OFF24}}

\begin{document}

\title{\Large\bf Assessing zero-shot generalisation behaviour in graph-neural-network interatomic potentials}

\author[1]{Chiheb Ben Mahmoud\thanks{chiheb.benmahmoud@chem.ox.ac.uk}}
\author[1]{Zakariya El-Machachi}
\author[1]{Krystian A. Gierczak}
\author[1]{John L. A. Gardner}
\author[1]{Volker L. Deringer}

\affil[1]{Inorganic Chemistry Laboratory, Department of Chemistry, University of Oxford, Oxford OX1 3QR, UK}

\date{}

\maketitle

\clearpage

\setstretch{1.5}

\begin{abstract}
With the rapidly growing availability of machine-learned interatomic potential (MLIP) models for chemistry, much current research focuses on the development of generally applicable and ``foundational'' MLIPs. 
An important question in this context is whether, and how well, such models can transfer from one application domain to another.
Here, we assess this transferability for an MLIP model at the interface of materials and molecular chemistry.
Specifically, we study \go{}, a model designed for the extended covalent network of graphene oxide, and quantify its zero-shot performance for small, isolated molecules and chemical reactions outside its direct scope---in direct comparison with a state-of-the-art model which has been trained in-domain.
Our work provides quantitative insight into the transfer and generalisation ability of graph-neural-network potentials and, more generally, makes a step towards the more widespread applicability of MLIPs in chemistry.
\end{abstract}

\section*{Introduction}

Machine-learned interatomic potentials (MLIPs) for atomistic simulations, trained on quantum-mechanical energy and force data, have advanced remarkably in recent years \cite{behler_first_2017, deringer_machine_2019, unke_machine_2021} and now almost routinely allow researchers to address a wide range of questions in chemistry and materials science~\cite{bartok_machine_2018, cheng_evidence_2020, zhou_device-scale_2023, zhang_modelling_2024}. Recently, MLIPs incorporating graph-based representations, commonly referred to as graph neural networks (GNNs)~\cite{batzner_e3-equivariant_2022, NEURIPS2022_4a36c3c5, batatia_design_2025, deng_chgnet_2023}, have emerged as cost-effective yet chemically rich models of atomic interactions. 
The favourable scaling of GNN-based MLIPs with the number of atomic species means that they are, in principle, able to cover elements from across the Periodic Table all in a single model~\cite{batatia_foundation_2023,deng_chgnet_2023,merchant_scaling_2023,yang_mattersim_2024}. 

The enhanced chemical versatility provided by GNNs has inspired the development of so-called ``pre-trained''~\cite{deng_chgnet_2023}, ``foundational''~\cite{batatia_foundation_2023}, or ``universal''~\cite{chen_universal_2022,yang_mattersim_2024} interatomic potentials. These models have been trained on large, structurally and chemically diverse datasets; they show promising baseline performance for a range of systems~\cite{focassio_performance_2024,ju_application_2025} and thus provide a practical tool for starting computational projects, as well as a basis for fine-tuning \cite{kaur_data-efficient_2025}. 
In the long run, one might want to employ these pre-trained MLIPs ``as is'', in a zero-shot manner, without additional training or adaptation. Zero-shot performance also yields an important indication of how well the underlying model generalises to unseen tasks and chemistries.
Understanding and improving the zero-shot behaviour of MLIPs is therefore an important challenge.

Herein, we study the zero-shot generalisation behaviour of \go{}  (Ref.~\citenum{elmachachi_accelerated_2024}), an MLIP model that was initially developed specifically for graphene oxide (GO). 
Conceptually, GO bridges the gap between pristine graphene and organic chemistry: its structural landscape involves a variety of bonding motifs from $\ce{sp}^2$ carbon sheets to oxygen-rich domains and reactive edge sites~\cite{dreyer_chemistry_2010}. 
We test whether this structural and chemical complexity may serve as a basis for transferability (albeit initially we thought of \go{} as a single- rather than general-purpose MLIP!),
subjecting \go{} to a range of out-of-domain benchmarks, from energetics to high-temperature molecular-dynamics (MD) simulations of chemical reactions. In this way, our present study explores: (i) the role of a chemically rich training dataset in building robust and generalisable MLIPs \cite{ben_mahmoud_data_2024}; (ii) the importance of GNN-based architectures in doing so; and (iii) the question whether \go{} could form a starting point for foundational MLIPs bridging materials and molecular chemistry.

\section*{Methodology}

\subsection*{The \go{} and \off{} models}

We focus on the \go{} model, which was built using the MACE architecture~\cite{NEURIPS2022_4a36c3c5, batatia_design_2025} together with a bespoke data-generation protocol~\cite{elmachachi_accelerated_2024}. 
Initial training data were generated ``from scratch'' using CASTEP+ML~\cite{stenczel_machine-learned_2023} (accelerating {\em ab initio} MD through on-the-fly fitting of GAP models\cite{bartok_gaussian_2010}), and then largely augmented through subsequent iterative training from MD trajectories driven by intermediate versions of MACE models. Over time, configurations with functionalised edges, involving hydroxyl (\ce{-OH}), aldehyde (\ce{-CHO}), and carboxylic acid (\ce{-CO2H}) moieties, were added to ensure good coverage of the structural and chemical features that might be expected to appear in a ``real-world'' GO sheet. Training labels, viz.\ total energies and forces, were obtained from density-functional-theory (DFT) computations performed with the plane-wave software CASTEP~\cite{clark_first_2005} using on-the-fly generated pseudopotentials and the Perdew--Burke--Ernzerhof (PBE) exchange--correlation functional~\cite{perdew_generalized_1996}. 

As a baseline for the current state-of-the-art (SOTA) in molecular modelling, we choose two variants of the \off{} family of MLIPs~\cite{kovacs_mace-off23_2023}: the ``large'' version of \offt{}, commonly referred to as \offt(L), which is trained on the SPICE dataset of molecular data version 1~\cite{eastman_spice_2023}, and \offf{}, which is trained on the SPICE dataset version 2~\cite{eastman_nutmeg_2024}. \offf{} is more similar to \go{} in terms of architecture, with the exception of the radial cut-off: 3.7~\AA for \go{} and 6.0~\AA for \offf{}. More details about the hyperparameters of all the GNNs used in this work are provided in the Supplementary Information. In the remainder of this work, we refer to \offt(L) simply as \offt.
In using \off{} models as benchmarks, it is important to note the different DFT levels of theory compared to \go{}: the SPICE labels were obtained from DFT computations with the $\omega$B97M-D3(BJ) exchange--correlation functional~\cite{mardirossian__2016,najibi_nonlocal_2018} and the def2-TZVPPD basis set~\cite{weigend_balanced_2005,rappoport_property-optimized_2010}.

\subsection*{Benchmark data}

\section*{Methodology}

\subsection*{The \go{} and \off{} models}

We focus on the \go{} model, which was built using the MACE architecture~\cite{NEURIPS2022_4a36c3c5, batatia_design_2025} together with a bespoke data-generation protocol~\cite{elmachachi_accelerated_2024}. 
Initial training data were generated ``from scratch'' using CASTEP+ML~\cite{stenczel_machine-learned_2023} (accelerating {\em ab initio} MD through on-the-fly fitting of GAP models\cite{bartok_gaussian_2010}), and then largely augmented through subsequent iterative training from MD trajectories driven by intermediate versions of MACE models. Over time, configurations with functionalised edges, involving hydroxyl (\ce{-OH}), aldehyde (\ce{-CHO}), and carboxylic acid (\ce{-CO2H}) moieties, were added to ensure good coverage of the structural and chemical features that might be expected to appear in a ``real-world'' GO sheet. Training labels, viz.\ total energies and forces, were obtained from density-functional-theory (DFT) computations performed with the plane-wave software CASTEP~\cite{clark_first_2005} using on-the-fly generated pseudopotentials and the Perdew--Burke--Ernzerhof (PBE) exchange--correlation functional~\cite{perdew_generalized_1996}. 

As a baseline for the current state-of-the-art (SOTA) in molecular modelling, we choose two variants of the \off{} family of MLIPs~\cite{kovacs_mace-off23_2023}: the ``large'' version of \offt{}, commonly referred to as \offt(L), which is trained on the SPICE dataset of molecular data version 1~\cite{eastman_spice_2023}, and \offf{}, which is trained on the SPICE dataset version 2~\cite{eastman_nutmeg_2024}. \offf{} is more similar to \go{} in terms of architecture, with the exception of the radial cut-off: 3.7~\AA for \go{} and 6.0~\AA for \offf{}. More details about the hyperparameters of all the GNNs used in this work are provided in the Supplementary Information. In the remainder of this work, we refer to \offt(L) simply as \offt.
In using \off{} models as benchmarks, it is important to note the different DFT levels of theory compared to \go{}: the SPICE labels were obtained from DFT computations with the $\omega$B97M-D3(BJ) exchange--correlation functional~\cite{mardirossian__2016,najibi_nonlocal_2018} and the def2-TZVPPD basis set~\cite{weigend_balanced_2005,rappoport_property-optimized_2010}.

\subsection*{Benchmark data}

We carry out initial tests using the revised version of the MD17 dataset (rMD17)~\cite{christensen_role_2020}. We select the 6 molecules from rMD17 that only contain the elements C, H, and O -- the only ones in the GO dataset, and thus the only ones that \go{} and other models directly fitted to its dataset can handle. For each molecule, we randomly select 1,000 configurations from the available trajectories. The rMD17 labels were obtained in the original work using the PBE functional and the def2-SVP basis set~\cite{perdew_generalized_1996,weigend_balanced_2005}.

The other test sets used in the present study are generated either by running MD simulations in the $NVT$ ensemble or by relaxing molecules. In both cases, we use \go{} to perform these tasks. We compute reference data using DFT, matching the settings for \go{} and \off, where applicable. For comparison to \go{}, labels are obtained from CASTEP by placing the molecules in large periodic cells ($> 20$~\AA). For \off, compatible labels are obtained using the Atomic Simulation Environment (ASE)~\cite{hjorth_larsen_atomic_2017} Python interface of Psi4~\cite{smith_p_2020}, version 1.4.

\begin{figure}
    \centering
    \includegraphics[]{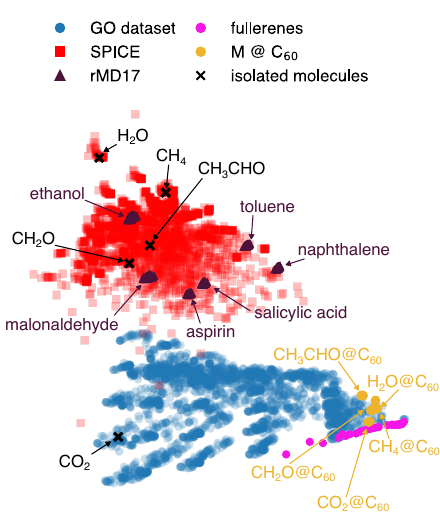}
    \caption{Visualising the structural and chemical space explored in the present study. We show a two-dimensional embedding of the MACE descriptor trained on the GO dataset~\cite{elmachachi_accelerated_2024}, using principal component analysis. The points of the map correspond to the training set of \go{} (blue), molecules containing C, H, and O atoms, representing $\approx 5$\% of the SPICE (version 1) dataset~\cite{eastman_spice_2023} (red), configurations from rMD17 trajectories~\cite{christensen_role_2020} (purple), a series of fullerenes with sizes ranging between 20 and 100 (magenta), five molecules encapsulated in \ce{C60} fullerene cages (yellow), and the same molecules in vacuum (black crosses).} 
    \label{fig:map}
\end{figure}

\subsection*{Data overlap between molecules and graphene oxide}

Before benchmarking \go{}, it is important to set performance expectations based on the similarity of the various test sets and the GO training set. In Fig.~\ref{fig:map}, we present a two-dimensional embedding, from principal component analysis (PCA), of the average atomistic features per snapshot as learned by \go{}. The use of average features eliminates the system-size dependence of the descriptors. 
We observe that static rMD17 molecules lie outside the scope of the training data ({\em blue}), but fall within the SPICE dataset domain ({\em red}), which constitutes the training data of \off. We should thus expect \off{} to outperform \go{} for static molecules. ``Acyclic'' molecules such as ethanol seem to be farther from the GO domain compared to cyclic molecules, such as aspirin. As a result, we expect \go{} to provide more accurate predictions for cyclic molecules compared to acyclic configurations.
Fullerenes ({\em magenta}) and encapsulated molecular species (``M @ \ce{C60}'', {\em yellow}) are located on the outskirts of the GO region of the map in Fig.~\ref{fig:map}---this is unexpected at first glance, as fullerenes are not part of the GO training data. However, some of their key characteristics can be learned from the GO backbone.

\section*{Zero-shot performance of \go{}}
In this section, we evaluate the performance of \go{} in predicting the energies and forces of small molecules, as well as vibrational spectra. 
Throughout this section, we use the terms ``error'' and ``root mean square error'' (RMSE) interchangeably. 

\subsection*{Numerical performance for MD17}
\begin{figure}
    \centering
    \includegraphics[]{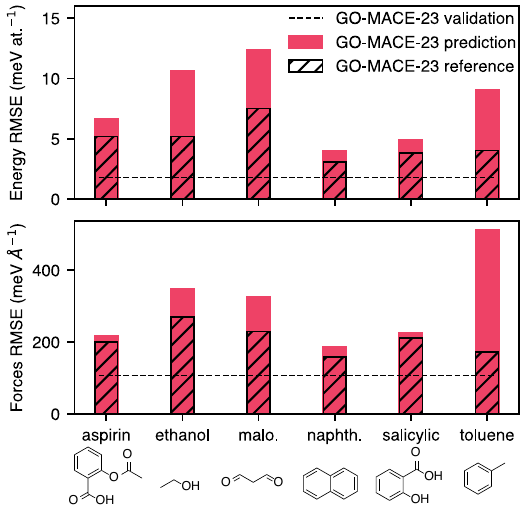}
    \caption{Energy and force errors on six trajectories from the revised MD17 dataset using \go{}. The bars represent the RMSE of quantities between \go{} predictions and rMD17 labels. The dashed area represents the errors between the DFT levels of theory used to label the GO dataset and the rMD17 dataset. The dashed line is the internal validation error of \go{}.} 
    \label{fig:go-rmd17}
\end{figure}

A common starting point in evaluating MLIP performance is in testing prediction errors for energies and forces. These tests can be more complex than they look at first glance, because their outcome will strongly depend on the type of data used for testing (see, e.g., Ref.~\citenum{thomas_du_toit_hyperparameter_2024}). In the present work, we are interested in zero-shot generalisability (without further modification of the model), which we here test by changing the application domain from extended GO structures to isolated (small) molecules.

We begin our series of zero-shot tests by evaluating the performance of \go{} for the relevant trajectories from the rMD17 dataset. In Fig.~\ref{fig:go-rmd17}, we summarise the prediction errors on total energies and atomic forces relative to the QM targets of the rMD17 molecules. Despite the differences in the levels of theory between \go{} and rMD17, we observe RMSE values below the often-quoted ``chemical accuracy'' of 1~kcal mol$^{-1}$ or $\approx$ 40~meV at.$^{-1}$.  
However, these errors can be significantly higher than the model's {\em internal} internal validation error for GO (1.8~meV~atom$^{-1}$ for energies and 109~meV~\AA$^{-1}$ for forces, shown as dashed lines in Fig.~\ref{fig:go-rmd17}), which is the case for malonaldehyde. The latter is an example of an acyclic molecule (not containing an aromatic ring) that is not well represented in the \go{} dataset.
To explore the origin of these errors, we performed DFT calculations, using the same parameters as used for training \go{}, on the different test snapshots, and we report the RMSE between the levels of theory, as shown by hatched bars in Fig.~\ref{fig:go-rmd17}. We find that \go{} is primarily constrained by its own training QM labels, as systematic discrepancy errors account for approximately 30\% to 90\% of the errors. 
For aspirin, naphthalene, and salicylic acid, \go{} introduces almost no additional errors beyond those inherent to its DFT training labels, and its prediction errors are comparable to its internal validation errors. \go{} introduces almost no additional errors to the predictions made on cyclic molecules, although the magnitude of the errors varies notably. 

This evaluation highlights the importance of contextualising zero-shot performance of pretrained ML models across datasets. Most of the force prediction errors for the rMD17 molecules stem from discrepancies in the underlying DFT, with the exception of toluene (which we address in the following). 
Figure \ref{fig:go-rmd17} suggests that molecules with structural motifs resembling those in a GO sheet are better captured by \go{}, reinforcing the importance of dataset choice for generalisability. 
While the ideal situation is to always compare data coming from uniform sources, we understand that this might not always be computationally feasible, underscoring the need for robust contextual analysis in ML model evaluations.

\subsection*{Toluene as a special case}
\begin{figure*}
    \centering
    \includegraphics[]{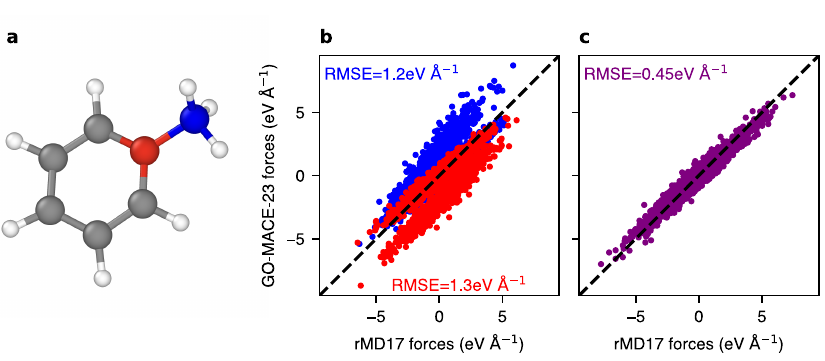}
    \caption{(a) Visualisation of a toluene molecule obtained using OVITO~\cite{stukowski_visualization_2010}. Red- and blue-coloured atoms are carbon atoms part of the aromatic ring and the attached methyl group, respectively. (b) Force components parity plot of the DFT-computed and \go{}-predicted forces for the carbon atoms labelled red and blue in panel (a). (c) Force parity plot of the sum of forces of the red- and blue- labelled carbon atoms.} 
    \label{fig:toluene}
\end{figure*}

\begin{figure}
    \centering
    \includegraphics[]{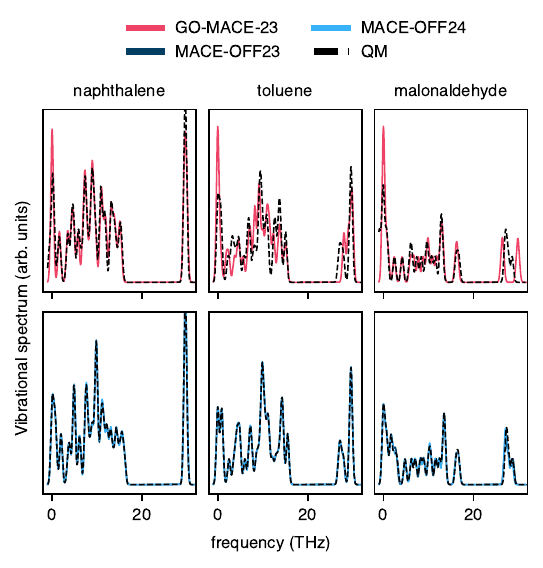}
    \caption{Molecular vibrational spectra computed with MLIPs ({\em solid lines}) and DFT (``QM'', {\em dashed lines}) for \go{}-relaxed naphthalene, toluene, and malonaldehyde molecules. The upper row characterises the out-of-domain performance of \go{} ({\em red}). The lower row shows the performance of SOTA MLIPs for molecules, viz.\ \off \cite{kovacs_mace-off23_2023} ({\em dark and light blue}). Note that the DFT data have been computed at the level corresponding to the training data of the respective MLIP model; the DFT data in the upper and lower rows are therefore slightly different.} 
    \label{fig:phonons}
\end{figure}

To better understand the performance limits of \go{}, we analyse the errors for toluene in more detail, as it exhibits the highest force prediction RMSE among all 6 rMD17 molecules considered here. Fig.~\ref{fig:toluene} summarises our approach to exploring possible sources of error. The toluene molecule contains an aromatic carbon atom directly bonded to an \ce{sp^3} carbon atom in a methyl group (\ce{-CH3}), coloured in red and blue in Fig.~\ref{fig:toluene}a, respectively. These two carbon atoms have the highest overall force errors exceeding 1.2~eV~\AA$^{-1}$ (Fig.~\ref{fig:toluene}b). The high force errors on these specific atoms indicate that \go{} is incapable of faithfully modelling their behaviour, due to the under-representation of similar atomic environments in the GO training set.

Most current MLIPs (including the MACE architecture) describe the total energy of a chemical system as a sum of atomic energies, following Refs.~\citenum{behler_generalized_2007} and \citenum{bartok_gaussian_2010}. While this decomposition is useful for training and extrapolating ML models, it is not inherently physical and has no direct counterpart in a quantum-mechanical computation: so it is possible for the MLIP to reproduce the global behaviour without capturing the expected {\em local} energy distribution. This issue is evident in the present case of toluene (Fig.~\ref{fig:toluene}c): the {\em combined} error for the sum of the forces is only one-third of the individual force-component errors.
The predicted atomic energies confirm this limitation (Fig.~S1): the ``red'' atom of the aromatic ring has the lowest predicted atomic energy of all the carbon atoms, while the ``blue'' atom of the methyl group has the highest. When averaging the energies of these two atoms, the methyl carbon and its direct neighbour have the lowest local energy across the randomly selected 200 snapshots in the trajectory (Fig. S1). More generally, further work is necessary to fully understand the local predictions of MLIPs, and steps towards this goal have been made\cite{chong_robustness_2023, chong_prediction_2025}.

\subsection*{Vibrational spectra}

The vibrational spectrum---which provides information about bending, twisting, and stretching of individual bonds---is a fingerprint of a molecule (and experimentally accessible), and is therefore an important test for an MLIP to accurately reproduce. 
To assess the ability of \go{} to predict vibrational spectra, we focus on three molecules from the rMD17 dataset: naphthalene and toluene representing the best and worst force predictions, respectively (cf.\ Fig.~\ref{fig:go-rmd17}), and malonaldehyde as an example of a molecule without a  6-membered aromatic ring (the principal structural fragment of graphene). We start by selecting a random snapshot from the three trajectories, then relax the molecules using \go. 
The force errors for the relaxed structures are 0.05~eV~\AA$^{-1}$ for naphthalene, 0.32~eV~\AA$^{-1}$ for toluene, and 0.22~eV~\AA$^{-1}$ for malonaldehyde. Then, we compute the spectra with the MLIP and DFT at the corresponding level, using finite displacements, from phonopy~\cite{togo_implementation_2023, togo_first-principles_2023}. We present the resulting vibrational spectra in the upper panels of Fig.~\ref{fig:phonons}. 
The \go-predicted spectra agree qualitatively with their DFT counterparts, and the quality of the prediction correlates well with the model's force accuracy. The low-frequency modes, in particular, are well reproduced, while the accuracy decreases for the high-frequency modes. 
A recent study in Ref.~\citenum{deng_systematic_2025} suggests that these discrepancies may arise from a softened potential-energy surface near the relevant snapshots, which could explain the reduced accuracy for high-frequency modes.

We compare \go{} to \offt{} and \offf{}, two SOTA molecular MLIP models trained on different versions of the SPICE molecular dataset (see Methodology section). We compute the vibrational spectra on the \go-relaxed molecules using \off{} and their corresponding DFT level of theory. The force errors of \offt{} are 0.003, 0.002, and 0.016~eV~\AA$^{-1}$ for naphthalene, toluene, and malonaldehyde, respectively. The force errors of \offf{} are 0.005, 0.003, and 0.005~eV~\AA$^{-1}$ for naphthalene, toluene, and malonaldehyde, respectively.  We report the spectra in the lower panels of Fig.~\ref{fig:phonons}. As shown in Fig.~\ref{fig:map}, the rMD17 molecules fall within the training domain of the \off{} models, which explains the models' high accuracy in predicting atomic forces. As a result, both \off{} models produce more accurate vibrational spectra, reproducing both high- and low-frequency modes. 

\subsection*{Fullerenes and encapsulated molecules}

\begin{figure}
    \centering
    \includegraphics[]{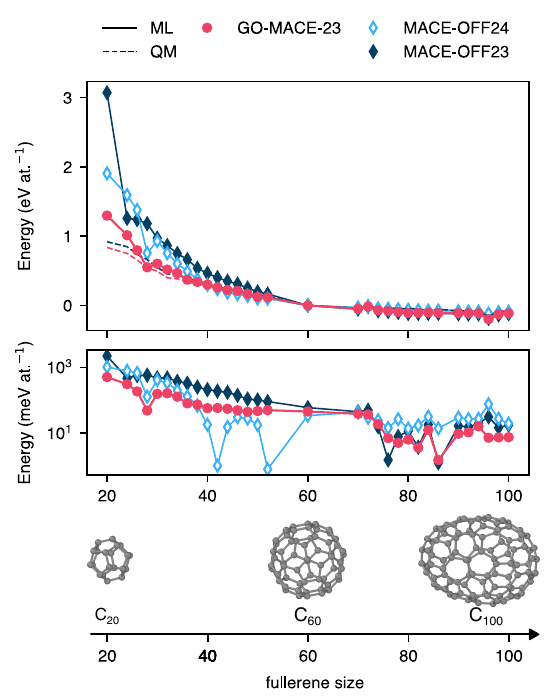}
    \caption{Evolution of the per-atom energy of fullerenes, obtained from Ref.~\citenum{barnard_fullerene_2023}, of sizes between 20 and 100 atoms computed with \go{} and its corresponding DFT level of theory ({\em red}), and \off{} and their corresponding DFT level of theory ({\em dark and light blue}). Similar to Fig.~\ref{fig:phonons}, lines represent the ML predictions, and the dashed lines represent the QM reference calculations. All energies are referenced to \ce{C60}. The lower panel describes the difference between energies computed with ML and QM, and expressed per atom. The rendered images show three fullerenes: \ce{C20}, \ce{C60}, and \ce{C100}.}
    \label{fig:fullerenes}
\end{figure}

\begin{figure}
    \centering
    \includegraphics[]{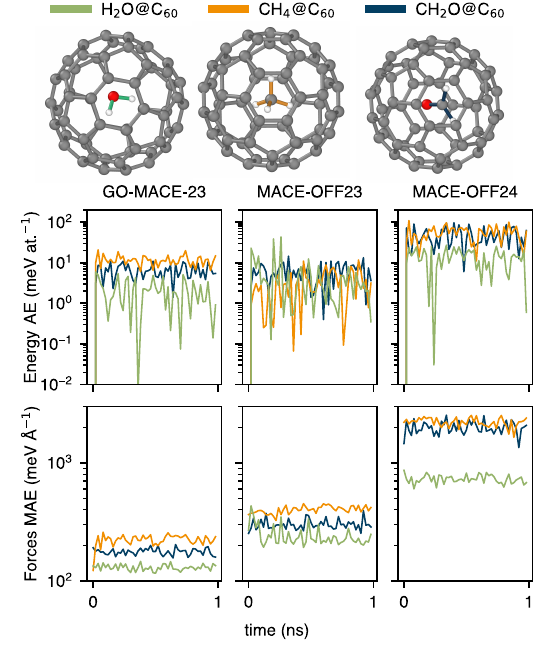}
    \caption{Evolution of energy and force RMSE between \go{} predictions and the corresponding DFT level of theory (left column), as well as between both \off{} variants and their respective DFT levels of theory (middle and right columns). The errors are calculated from 1~ns trajectories at 500 K for \ce{H2O}, \ce{CH4}, and \ce{CH2O} enclosed in a \ce{C60} fullerene. The trajectories are driven by \go{}.}
    \label{fig:fullerenes-anneals}
\end{figure}

We use a series of fullerene molecules as another benchmark to quantify the transferability of \go{} (and \off). The smallest fullerene is \ce{C20}, containing only five-membered rings of carbon atoms and no six-membered ones. Consequently, its curvature is large, and the fullerene is found to be the most stable \ce{C20} conformer using MP2 calculations~\cite{parasuk_c20_1991}. 
Larger fullerenes are energetically and structurally closer to graphene and graphite, and should therefore be closer to the training domain of \go{} (cf.\ Fig.~\ref{fig:map}).

Both \go{} and \off{} variants reproduce the general trend of growing stabilisation with fullerene size, as shown in Fig.~\ref{fig:fullerenes}. Prediction errors are highest for the smaller fullerenes, with RMSE values higher than $>100$~meV~at.$^{-1}$, likely due to their high curvature. For \ce{C60}, the RMSE decreases to around $50$~meV~at.$^{-1}$ for all MLIPs. For small fullerenes ($< 60$ carbon atoms), \go{} performs better than both \off models: we presume that this is due to the fact that it has encountered some curved graphene sheets, including various odd-membered rings, during training. Note, however, that the latter are only a small fraction of the training data: the ring-size distribution in the \go{} dataset is 1:600 for 5:6-membered rings. \offf{} significantly outperforms both \go{} and \offt{} for fullerenes with the sizes of 42 and 50 atoms, hinting towards the existence of relevant motifs within the updated version of the SPICE dataset. This requires further investigation.

In a recent study, Vyas at al.~showed how formaldehyde (\ce{CH2O}) can be inserted into a \ce{C60} molecule by subsequent organic reaction steps~\cite{vyas_squeezing_2024}, expanding on existing work on endohedral fullerenes~\cite{popov_endohedral_2013,bloodworth_synthesis_2022}. In the context of the present work, we show in Fig.~\ref{fig:fullerenes-anneals} three case studies that have been discussed in the literature: encapsulated water (written as ``\ce{H2O}@\ce{C60}'')~\cite{carrillo-bohorquez_encapsulation_2021}, encapsulated methane (``\ce{CH4}@\ce{C60}'')~\cite{bloodworth_first_2019}, and encapsulated formaldehyde (``\ce{CH2O}@\ce{C60}'')~\cite{vyas_squeezing_2024}. 

We use \go{} to drive long MD trajectories of the three species in the $NVT$ ensemble at $T=500$~K, for 1~ns with a 0.5~fs timestep. Such simulations can be challenging test cases \cite{stocker_how_2022}, especially given the fusion temperature of \ce{C60} is estimated to be around 550~K~\cite{elvers_ullmanns_2011}. 
We re-label snapshots from these MD trajectories with \go{} and its corresponding DFT method, as well as \off{} and its corresponding DFT method. In Fig.~\ref{fig:fullerenes-anneals}, we show the errors, expressed as absolute errors (AE) for energies and mean absolute errors (MAE) for forces rather than our usual RMSE, for snapshots sampled every 20~ps. 
Both MLIPs exhibit similar energy prediction errors, with \go{} performing better for the larger encapsulated molecules, and \offt{} for \ce{H2O}@\ce{C60}. However, \go{} consistently yields lower force prediction errors across all of the test cases.
This poorer performance of \offt{} and \offf{} may be attributed to the fact that fullerenes and encapsulated molecules are not present within the two versions of the SPICE training set. Additionally, \go{} has encountered small molecules, such as \ce{CO} and \ce{H2O}, near GO surfaces in its training data. Also, it is possible that \go{} is accessing regions of configurational space that would be deemed unphysical by \off. Uncertainty estimation of predictions made by these models could provide an answer, even partially, to this question.

In the Supplementary Information, we show two additional cases of encapsulated molecules, carbon dioxide and acetaldehyde, the heavier homologue of \ce{CH2O}. Acetaldehyde is a challenging test case for \go{}, and has most likely not been seen during training (cf. Fig.~\ref{fig:map}). It is a thought experiment, of course, for the time being.

\section*{Experiments}

Beyond the zero-shot performance evaluation so far, we carry out additional numerical experiments. These explore aspects of MLIP fitting methodology and provide an initial test for descriptions of gas-phase fragmentation reactions.
 
\subsection*{Model choice (I): Effect of equivariant messages}

\begin{table}
\small
  \caption{\ Energy and force prediction RMSE as a function of the maximum rank of the equivariant hidden messages in the MACE architecture for trajectories from the rMD17 dataset. Errors are computed with respect to the DFT level of theory of rMD17. The lowest RMSE values for each molecules are highlighted in bold}\label{tab:equivar}
  \begin{tabular*}{\linewidth}{@{\extracolsep{\fill}}lccccccc}
    \hline
    & \multicolumn{3}{c}{Energy RMSE (meV~at.$^{-1}$)} & & \multicolumn{3}{c}{Force RMSE (eV~\AA$^{-1}$)} \\
    \hline

    $\texttt{max L}$& 0 & 1 & 2 & & 0 & 1 & 2\\
    \hline
    aspirin &  6.2 & 6.6 & {\bf 4.9} & & 0.25 & {\bf 0.22} & 0.28\\
    ethanol &  12.3 & {\bf 10.6} & 12.2 & & 0.49 & {\bf 0.35} & 0.48\\
    malonaldehyde &  {\bf 7.7} & 12.3 & 9.2 & & 0.28 & 0.33 & {\bf 0.25}\\
    naphthalene &  {\bf 3.3} & 4.0 & 3.6 & & 0.18 & 0.18 & {\bf 0.17}\\
    salicylic acid &  5.3 & {\bf 4.9} & 6.8 & & {\bf 0.22} & {\bf 0.22} & 0.26\\
    toluene &  {\bf 5.6} & 9.1 & 6.9 & & 0.32 & 0.51 & {\bf 0.25}\\
    \hline
  \end{tabular*}
\end{table}

The MACE architecture underlying \go{} incorporates both invariant hidden features and equivariant hidden features of rank $\texttt{L}=1$. To test the role of equivariance, we trained two modified versions of the model by varying MACE's internal symmetry rank. Specifically, we trained an invariant model by setting the highest rank of the internal features to $\texttt{max L}=0$, and a higher-order equivariant model by setting $\texttt{max L}=2$. This allows us to explore the possible correlation between the physical symmetries of an MLIP and its out-of-domain performance.

In Table \ref{tab:equivar}, we compare the performance of MACE models using invariant {\em vs} equivariant messages with different maximum rank $\texttt{max L}\in \{0,~1,~2\}$. We calculate the prediction errors for all relevant rMD17 molecules, using MACE models re-fitted with different $\texttt{max L}$ values. 
We notice that the original \go{} model ($\texttt{max L}=1$) does not systematically outperform its invariant counterpart ($\texttt{max L}=0$). For example, the invariant model yields better energy predictions for toluene, aspirin, and naphthalene, as well as better force predictions for salicylic acid, compared to \go{}. A similar trend is observed when comparing \go{} to the $\texttt{max L}=2$ MACE model. Regardless of the benchmark reference calculation, we observe no clear correlation between $\texttt{max L}$ and model performance, suggesting that equivariance and symmetry preservation play a limited role in generalisation for these domains. 
A particularly notable case is the toluene trajectory, where \go{} is the {\em worst}-performing model of the three, in terms of total energy and force predictions (cf.\ Fig.~\ref{fig:toluene}).

\subsection*{Model choice (II): Other GNN architectures}
\begin{table}
\small
\setlength{\tabcolsep}{1pt}
  \caption{\ Energy and force prediction RMSE different GNN architectures trained on the GO dataset for trajectories from the revised MD17 dataset. Errors are computed with respect to the DFT level of theory of rMD17. The lowest RMSE values for each molecules are highlighted in bold}\label{tab:other-models}
  \begin{tabular*}{\linewidth}{@{\extracolsep{\fill}}lccccc}
    \hline
    & \multicolumn{4}{c}{Energy RMSE (meV~at.$^{-1}$)} \\
    \hline
    & \go{} & SchNet & TensorNet & NequIP & PaiNN\\
    \hline
aspirin &  6.6 & 22.4 & 6.6 & \textbf{5.7} & 11.3 \\
ethanol &  \textbf{10.6} & 33.4 & 17.4 & 17.2 & 27.6 \\
malonaldehyde &  12.3 & 38.5 & 10.8 & \textbf{8.8} & 17.2 \\
naphthalene &  4.0 & 9.9 & 5.0 & \textbf{3.9} & 5.8 \\
salicylic &  4.9 & 19.7 & 5.6 & \textbf{3.9} & 7.4 \\
toluene &  9.1 & 16.8 & \textbf{8.7} & 24.0 & 14.0 \\
    \hline
    & \multicolumn{4}{c}{Force RMSE (eV~\AA$^{-1}$)} \\
    \hline
    & \go{} & SchNet & TensorNet & NequIP & PaiNN \\
    \hline
aspirin& \textbf{0.22} & 0.86 & 0.38 & 0.31 & 0.57\\
ethanol& \textbf{0.35} & 1.13 & 0.61 & 0.47 & 1.01\\
malonaldehyde& \textbf{0.33} & 0.98 & 0.34 & \textbf{0.33} & 0.38\\
naphthalene& \textbf{0.18} & 0.54 & 0.21 & 0.21 & 0.30\\
salicylic& 0.22 & 0.42 & 0.24 & \textbf{0.19} & 0.25\\
toluene& 0.51 & 0.59 & \textbf{0.28} & 0.38 & 0.32\\
\hline
  \end{tabular*}
\end{table}

To further investigate the effect of design choices made for several popular GNNs on their generalisability, we trained multiple models on the \go{} training dataset, using the universal interface {\tt graph-pes}~\cite{gardner_graph_2025}. Particularly, we used the SchNet~\cite{schutt_schnetpack_2019}, PaiNN~\cite{schutt_equivariant_2021}, TensorNet~\cite{simeon_tensornet_2023}, and NequIP~\cite{batzner_e3-equivariant_2022} architectures. Details about hyperparameters and validation errors on the \go{} dataset are provided in the Supplementary Information.

Table \ref{tab:other-models} shows that \go{}, TensorNet, and NequIP generally yield low RMSE on most molecules for both energy and force predictions. For instance, NequIP achieves low energy errors on aspirin and malonaldehyde, whereas TensorNet performs best for toluene. Meanwhile, \go{} has the best errors in force predictions for ethanol and naphthalene. 
These variations demonstrate that even closely related equivariant models can extract distinct mappings from the same data, influenced by subtle differences in model design and hyperparameters.

These results highlight the importance of the MLIP architecture in capturing relevant atomistic information and transferring it beyond the training set. The extrapolation is not trivial and depends not only on the quality of the training data or the fit but also on the architecture itself. 
Notably, as shown in the Supplementary Information, \go{} has the lowest energy validation errors on the GO dataset, yet NequIP outperforms it for several rMD17 molecules. These results underscore the need for systematic out-of-domain validation to fully assess model generalisation.

\begin{figure}
    \centering
    \includegraphics[]{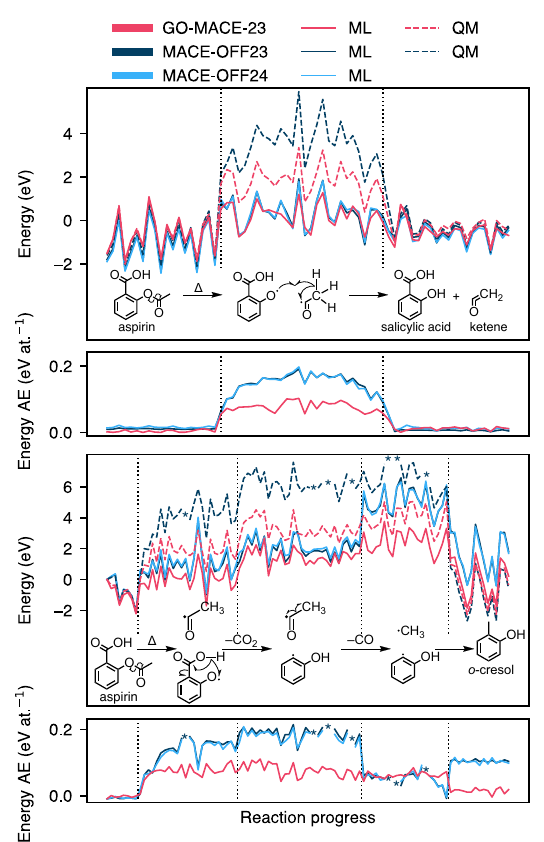}
    \caption{Energy profiles of two exemplary high-temperature molecular-dynamics simulations computed with \go{}, \offt{}, \offf{}, and their respective QM references. The MD trajectories are driven by \go{} and maintained at 1,500~K. The first panel describes a reaction pathway to produce salicylic acid and ketene (\ce{H2CCO}) from aspirin. The third panel describes the decomposition of aspirin through a series of decarbonylations and decarboxylations to produce $o$-cresol. Second and fourth panels describe the difference between energies computed with ML and QM, for the first and second reactions, respectively, and expressed per atom. The asterisks correspond to failed DFT calculations after 30 self-consistent cycles.}
    \label{fig:decarbolynation}
\end{figure}
\subsection*{Transferability to chemical reactions}
The long-term goal of molecular interatomic potentials is to describe entire reaction mechanisms, rather than just the reactants and products. MLIPs are increasingly being used to describe transition states of reactions in vacuum~\cite{komp_low-cost_2022,choi_prediction_2023} and in explicit solvent~\cite{zhang_modelling_2024}. 
While \go{} has been trained on various rearrangements, decarbonylation reactions, etc., it has not been explicitly trained on molecular reaction mechanisms. This makes it a particularly challenging and relevant ``real-world'' benchmark for complex chemical transformations. 

We use \go{} to run a series of MD trajectories of an aspirin molecule in a periodic simulation cell of 30~\AA$^3$, using the $NVT$ ensemble at $T=1,500$~K. We also re-label the trajectories using the DFT reference method of \go{}, as well as using both \off{} variants and their DFT reference method. 
In Fig.~\ref{fig:decarbolynation}, we report two reaction pathways demonstrating the thermally driven decomposition of aspirin in vacuum into radical species which then recombine forming different molecules.

The upper panels of Fig.~\ref{fig:decarbolynation} depict the formation of reactive ketene and salicylic acid, a process involving the breaking of an ester bond. 
The reverse reaction was first described in Ref.~\citenum{nightingale_method_1926}. Both \go{} and the \off{} variants accurately capture the energetics of the reactants and products. However, they significantly underestimate the energy of the intermediates. Despite this underestimation, the predicted average energy of the intermediates remains higher than that of the more stable reactants or products.
This poor performance of both MLIPs is expected, as they are not explicitly trained on reaction pathways, and their underlying datasets do not include radicals or ions. In addition, these MLIPs were not able to reproduce the energy of the isolated radicals. 
Stocker et al.~\cite{stocker_machine_2020} have previously discussed the limitations of MLIPs in accurately describing chemical reactions when radicals are not explicitly incorporated in the training data.

The lower panels of Fig.~\ref{fig:decarbolynation} illustrate the formation of an $o$-cresol molecule through a series of decarboxylation and decarbonylation steps. This reaction pathway shares the first set of radicals with the upper panel, with similar geometries, before developing into a different pathway. As with the previous pathway, all tested MLIPs underestimate the energy of the intermediate steps. 
The two models from the \off{} family in particular overestimate the energy of the product system. 

\section*{Conclusions}
Located at the interface of materials and molecular modelling, graphene oxide offers an opportunity to connect these different domains of atomistic machine learning. 
In the present work, we have systematically assessed the zero-shot transferability of \go{}, an MLIP trained on data for GO, across relevant chemical benchmarks. We found good---perhaps surprisingly good---zero-shot performance compared to \off, a pre-trained model for molecular chemistry. The accuracy of both models decreases when describing reaction pathways, especially when radical species are involved. 

Our study has tested the behaviour of recently proposed GNN MLIP models and their transferability, and we think that it can have implications for the future development of ``foundational'' models for atomistic simulations. 
Our results emphasise that including chemical reactivity in the training data is important in finding reaction pathways: in the process of building the \go{} model~\cite{elmachachi_accelerated_2024}, we have sampled this reactivity in high-temperature MD simulations, and a similar approaches have been taken, e.g., for the bulk carbon--hydrogen~\cite{ibragimova_unifying_2025}, carbon--oxygen~\cite{zarrouk_experiment-driven_2024}, and organic condensed~\cite{zhang_exploring_2024} systems. 
We think that local-environment diversity will be as important as the chemical space coverage (e.g., the number of chemical species) in defining foundational models -- this might include the addition of radical species (cf.\ Fig.~\ref{fig:decarbolynation}) to the training data, either through very-high-temperature MD exploration or perhaps by explicitly involving ``broken'' bonds in the training protocol. Steps in this direction have been reported very recently~\cite{zhang_ani-1xbb_2025}. 

Despite its limitation to the three elements C, H, and O, the \go{} model seems to provide a suitable starting point to study a wider range of chemistry-related questions than it was initially intended for, and we view this as a highly encouraging finding. 
We believe that together with improved data-generation strategies \cite{ben_mahmoud_data_2024} as well as suitable workflows and automation approaches~\cite{podryabinkin_active_2017, young_transferable_2021, van_der_oord_hyperactive_2023, liu_automated_2024}, truly universal MLIPs for molecular systems, and for extended material structures built up from them, are coming within reach.

\clearpage

\section*{Author contributions}
C.B.M., Z.E.-M., and V.L.D. designed the research. K.A.G. carried out pilot studies, and C.B.M. and Z.E.-M. carried out the final numerical experiments. J.L.A.G. provided code and methodology for MLIP fitting. All authors contributed to discussions. C.B.M. and V.L.D. wrote the manuscript, and all authors reviewed and approved the final version.

\section*{Data availability}

Data supporting the present study are available at \href{https://github.com/cbenmahm/GO-Zero-Shot}{https://github.com/cbenmahm/GO-Zero-Shot}. An archived version of GraphPES is available at \href{https://zenodo.org/records/14956211}{https://zenodo.org/records/14956211}

\section*{Acknowledgements}
We thank J.~Holownia for useful discussions, and F.~Duarte for useful discussions and comments on the manuscript.
C.B.M. acknowledges funding from the Swiss National Science Foundation (SNSF) under grant number 217837.
We are grateful for support from the EPSRC Centre for Doctoral Training in Theory and Modelling in Chemical Sciences (TMCS), under grant EP/L015722/1 (Z.E.M.).
J.L.A.G. acknowledges a UKRI Linacre - The EPA Cephalosporin Scholarship, support from an EPSRC DTP award [grant number EP/T517811/1], and from the Department of Chemistry, University of Oxford.
V.L.D. acknowledges support from the Engineering and Physical Sciences Research Council [grant number EP/V049178/1] and UK Research and Innovation [grant number EP/X016188/1]. 
We are grateful for computational support from the UK national high performance computing service, ARCHER2, for which access was obtained via the UKCP consortium and funded by EPSRC grant ref EP/X035891/1.

\section*{References}
\vspace{2mm}

\bibliographystyle{rsc} 
\providecommand*{\mcitethebibliography}{\thebibliography}
\csname @ifundefined\endcsname{endmcitethebibliography}
{\let\endmcitethebibliography\endthebibliography}{}

\end{document}